# Dirac surface states in intrinsic magnetic topological insulators EuSn$_2$As$_2$ and MnBi$_{2n}$Te$_{3n+1}$

Hang Li,[1,2,#] Shun-Ye Gao,[1,2,#] Shao-Feng Duan,[3,#] Yuan-Feng Xu,[1,4,#] Ke-Jia Zhu,[1,2,#] Shang-Jie Tian,[5,#] Jia-Cheng Gao,[1,2] Wen-Hui Fan,[1,2] Zhi-Cheng Rao,[1,2] Jie-Rui Huang,[1,2] Jia-Jun Li,[1,2] Da-Yu Yan,[1,2] Zheng-Tai Liu,[6] Wan-Ling Liu,[6] Yao-Bo Huang,[7] Yu-Liang Li,[8] Yi Liu,[8] Guo-Bin Zhang,[8] Peng Zhang,[9] Takeshi Kondo,[9,10], Shik Shin,[9,10], He-Chang Lei,[5] You-Guo Shi,[1,11] Wen-Tao Zhang,[3,12,*] Hong-Ming Weng,[1,2,11,*] Tian Qian,[1,11,*] Hong Ding[1,2,11,13]

[1] *Beijing National Laboratory for Condensed Matter Physics and Institute of Physics, Chinese Academy of Sciences, Beijing 100190, China*
[2] *University of Chinese Academy of Sciences, Beijing 100049, China*
[3] *Key Laboratory of Artificial Structures and Quantum Control (Ministry of Education), School of Physics and Astronomy, Shanghai Jiao Tong University, Shanghai 200240, China*
[4] *Max Planck Institute of Microstructure Physics, Halle 06120, Germany*
[5] *Department of Physics and Beijing Key Laboratory of Opto-electronic Functional Materials and Micro–nano Devices, Renmin University of China, Beijing 100872, China*
[6] *State Key Laboratory of Functional Materials for Informatics, Shanghai Institute of Microsystem and Information Technology (SIMIT), Chinese Academy of Sciences, Shanghai 200050, China*
[7] *Shanghai Synchrotron Radiation Facility, Shanghai Advanced Research Institute, Chinese Academy of Sciences, 201204 Shanghai, China*
[8] *National Synchrotron Radiation Laboratory, University of Science and Technology of China, Hefei 230029, China*
[9] *Institute for Solid State Physics, University of Tokyo, Kashiwa, Chiba 277-8581, Japan.*
[10] *AIST-UTokyo Advanced Operando-Measurement Technology Open Innovation Laboratory (OPERANDO-OIL), Kashiwa, Chiba 277-8581, Japan*
[11] *Songshan Lake Materials Laboratory, Dongguan, Guangdong 523808, China*
[12] *Collaborative Innovation Center of Advanced Microstructures, Nanjing 210093, China*
[13] *CAS Center for Excellence in Topological Quantum Computation, University of Chinese Academy of Sciences, Beijing 100049, China*

[#] These authors contributed equally to this work.
[*] Corresponding authors: tqian@iphy.ac.cn, hmweng@iphy.ac.cn, wentaozhang@sjtu.edu.cn

**Abstract**

In magnetic topological insulators (TIs), the interplay between magnetic order and nontrivial topology can induce fascinating topological quantum phenomena, such as the quantum anomalous Hall effect, chiral Majorana fermions and axion electrodynamics. Recently, a great deal of attention has been focused on the intrinsic magnetic TIs, where disorder effects can be eliminated to a large extent, which is expected to facilitate the emergence of topological quantum phenomena. In despite of intensive efforts, experimental evidence of the topological surface states (SSs) remains elusive. Here, by combining first-principles calculations and angle-resolved photoemission spectroscopy (ARPES) experiments, we have revealed that $EuSn_2As_2$ is an antiferromagnetic TI with observation of Dirac SSs consistent with our prediction. We also observe nearly gapless Dirac SSs in antiferromagnetic TIs $MnBi_{2n}Te_{3n+1}$ ($n$ = 1 and 2), which were absent in previous ARPES results. These results provide clear evidence for nontrivial topology of these intrinsic magnetic TIs. Furthermore, we find that the topological SSs show no observable changes across the magnetic transition within the experimental resolution, indicating that the magnetic order has quite small effect on the topological SSs, which can be attributed to weak hybridization between the localized magnetic moments, from either 4$f$ or 3$d$ orbitals, and the topological electronic states. This provides insights for further research that the correlations between magnetism and topological states need to be strengthened to induce larger gaps in the topological SSs, which will facilitate the realization of topological quantum phenomena at higher temperatures.

Time-reversal symmetry has played a key role in topological quantum states of matter. The earliest discovered topological insulator (TI), Chern insulator with the integer quantum Hall effect, requires breaking the time-reversal symmetry [1-3]. The thinking and research on time-reversal symmetry in condensed matter systems directly led to the discovery of time-reversal-invariant (TRI) $Z_2$ TIs with the quantum spin Hall effect [4-7]. The introduction of magnetism into the $Z_2$ TIs can produce more exotic topological quantum phenomena, such as the quantum anomalous Hall effect [8-14], axion insulator states [15-21] and chiral Majorana fermions [22].

The quantum anomalous Hall effect has been first realized in magnetically doped $(Bi,Sb)_2Te_3$ thin films [9,10]. In the magnetically doped TIs, the magnetic impurities usually introduce strong inhomogeneity, which is believed to be one of the main reasons that the quantum anomalous Hall effect usually appears at extremely low temperatures (< 100 mK), hindering further exploration of topological quantum effects. A direct solution to avoid disorder is to seek for intrinsic magnetic TIs, which have magnetic order in the stoichiometric compositions.

In the past year, significant progress has been made in this field [16-21,23-39]. Theory has predicted several intrinsic antiferromagnetic (AF) TIs, such as $MnBi_2Te_4$ [16,19-21,33,40] and $EuIn_2As_2$ [17], while most experimental studies focused on $MnBi_2Te_4$ [18,21,25-27,30-39]. Very recently, a quantized Hall plateau at $h/e^2$ was realized in few-layer $MnBi_2Te_4$ sheets under moderate magnetic fields of several Tesla [30,32]. The intriguing phenomenon was attributed to the transition from an AF TI to a Chern insulator driven by magnetic fields. Angle-resolved photoemission spectroscopy (ARPES) experiments on $MnBi_2Te_4$ and $MnBi_4Te_7$ revealed a large gap of ~ 100 meV in the AF phase, which was considered to be associated with the time-reversal-symmetry breaking [21,25-29]. However, it is confusing that the large gap remains when the time-reversal symmetry is restored at temperatures much higher than the AF transition temperature ($T_N$) [21,25-29]. Another ARPES experiment on $EuSn_2P_2$ did not obtain information about the topological surface states (SSs), because the samples were hole doped with the bulk band gap above the Fermi level ($E_F$) [24].

In this work, we not only reveal the Dirac SSs of $MnBi_2Te_4$ and $MnBi_4Te_7$ within the large gap observed in previous ARPES studies, but also discover another intrinsic magnetic TI $EuSn_2As_2$ by combining first-principles calculations and ARPES measurements. Through a systematic study of these magnetic TIs, we obtain a comprehensive picture that while the magnetic order plays an important role in time-reversal-symmetry breaking, the coupling strength between the local magnetic moments and the topological electronic states is critical for the size of the opened gap in the topological SSs.

EuSn$_2$As$_2$ has a layered crystal structure with space group *R*-3*m* [Fig. 1(a)]. Each trigonal Eu layer is sandwiched between two buckled honeycomb SnAs layers. Two adjacent SnAs layers are coupled by van der Waals force. This allows EuSn$_2$As$_2$ to be easily exfoliated into few-layer sheets like MnBi$_2$Te$_4$ [26,30-32,41]. A previous study has revealed that EuSn$_2$As$_2$ undergoes a transition from a paramagnetic (PM) phase to an AF phase around 25 K [41], which is consistent with our measurements in Figs. 1(b) and 1(c). In the AF phase, the Eu 4*f* magnetic moments form an *A*-type AF structure, *i.e.*, ferromagnetic *a-b* planes coupled antiferromagnetically along the *c* axis. In addition, when the magnetic fields are perpendicular to the *c* axis, the susceptibility $\chi(T)$ in Fig. 1(b) shows an upturn below 10 K and the isothermal magnetization *M*(*H*) at 2 K in Fig. 1(c) increases rapidly at low fields, indicating an in-plane ferromagnetic component probably due to canting of the magnetic moments.

We first analyzed the topological properties of EuSn$_2$As$_2$ in the PM phase. We treated the Eu 4*f* states as core states in the pseudopotential band calculations of the PM phase in Fig. 1(e), since the localized Eu 4*f* states are located at ~ 1.7 eV below $E_F$ [Fig. 2(a)] and have negligible coupling with the other bands near $E_F$. The calculated bands show a continuous gap throughout the Brillouin zone (BZ) near $E_F$. For three-dimensional insulators with both inversion and time-reversal symmetries, one can use the Fu-Kane formula to compute the $Z_2$ invariant based on the parity eigenvalues at the eight TRI points in the BZ [42]. The numbers of occupied bands of odd parity at the eight TRI points are listed in Table I. The obtained invariant $Z_2 = 1$ indicates that EuSn$_2$As$_2$ is a strong TI in the PM phase.

In the band calculations of the AF phase in Fig. 1(f), we consider two metastable AF phases with the magnetic moments along the *b* (AF-*b*) and *c* (AF-*c*) axes, respectively. The Hubbard interaction *U* on the Eu 4*f* electrons was set to be 5 eV, in order to make the energy position of the Eu 4*f* bands consistent with the experimental results in Fig. 2(a). In the AF configuration, the pseudopotential of Eu must treat the 4*f* states as valence states. The calculated bands of the two AF phases are almost identical. The continuous gap throughout the BZ remains in the AF phases. In both AF phases of EuSn$_2$As$_2$, the time-reversal symmetry is broken and the inversion symmetry is preserved. One can compute the Z$_4$ invariant based on the parity eigenvalues at the eight TRI points. The numbers of occupied bands of odd parity at the eight TRI points are listed in Table I. The obtained invariant Z$_4$ = 2 indicates that EuSn$_2$As$_2$ is an axion insulator in the AF phase regardless of the spin orientations.

Based on the above analysis, EuSn$_2$As$_2$ transforms from a strong TI in the PM phase to an axion insulator in the AF phase below $T_N$. The effects of the topological phase transition on the Dirac SSs depend on the magnetic structures, spin orientations, and sample surfaces. Here, we consider the SSs on the (001) surface in the AF-$b$ and AF-$c$ phases. The analysis is similar to that for EuIn$_2$As$_2$ [17]. In both phases, the $\mathcal{T}\ell_{1/2}$ symmetry is broken at the (001) surface, where $\mathcal{T}$ is the time-reversal symmetry and $\ell_{1/2}$ is a translation operation of half of the magnetic unit cell along the $c$ axis, as indicated in Fig. 1(a). The breaking of $\mathcal{T}\ell_{1/2}$ symmetry causes the Dirac SSs to open an energy gap at $\bar{\Gamma}$. However, in the AF-$b$ phase, the vertical mirror symmetry ($M_y$) perpendicular to the magnetic moments is preserved and can protect gapless Dirac SSs away from $\bar{\Gamma}$ on the (001) surface. Our calculations in Fig. 1(g) show that the $k_y = 0$ plane has a nonzero mirror Chern number $n_{M_y=\pm i} = \mp 1$, indicating that EuSn$_2$As$_2$ is also a topological crystalline insulator in the AF-$b$ phase. The nonzero mirror Chern number protects a gapless Dirac cone on the $k_y = 0$ line on the (001) surface. As seen in Fig. 1(h), the calculated Dirac point is located 0.003 Å$^{-1}$ deviating from $\bar{\Gamma}$ on the $k_y = 0$ line. In the AF-$c$ phase, all vertical mirror symmetries are broken since the magnetic moments are parallel to them, leaving gapped Dirac SSs on the (001) surface. Note that the opened gap at $\bar{\Gamma}$ is too small (< 1 meV) to be resolved in the calculations in Fig. 1(i). The above analysis is illustrated in the schematic diagram in Fig. 1(j).

We then investigated the electronic structures on the (001) surface of EuSn$_2$As$_2$ with ARPES measurements. The electronic structures measured at different photon energies ($hv$) in Fig. 2(b) have no obvious changes. We use the data collected with $hv$ = 29 eV to illustrate the electronic structures near $E_F$ in the PM phase in Figs. 2(c)-2(e). The data reveal that all near-$E_F$ bands lie around the BZ center $\bar{\Gamma}$. Two hole-like bands (labelled as α and β) form two circular FSs centered at $\bar{\Gamma}$. In addition, one can see a small feature at $E_F$ at $\bar{\Gamma}$ (labelled as γ), which should be the bottom of an electron-like band, and an "M"-shaped band below $E_F$ (labelled as δ). The experimental data are consistent with the calculated valence bands in Fig. 2(f) except for a rigid band shift of ~ 0.18 eV. This suggests that the EuSn$_2$As$_2$ samples are hole doped, which is similar to the case of EuSn$_2$P$_2$ [24].

For the hole-doped samples, conventional ARPES measurements cannot obtain the information in the band gap. Instead, we used time-resolved ARPES (tr-ARPES) with the pump-probe method to measure the unoccupied electronic states above $E_F$. In Fig. 3(a), the snapshots of ARPES intensity at different pump-probe delay times reveal Dirac-like band dispersions at ~ 0.4 eV above $E_F$. To illustrate the topological properties of the Dirac band, we combine the tr-ARPES data with the conventional ARPES data in Fig. 3(d). The lower branch of the Dirac band connects to the band α. The bands β and γ connect just above $E_F$ and constitute a single band with an "M" shape. The upper branch of the Dirac band connects to the electron-like conduction band. It is expected that the development of long-range AF order with an *A*-type magnetic structure leads to band folding along $k_z$ and band reconstruction. We do not identify the effects in the bulk electronic structures because of low spectral intensity of the tr-ARPES data. The stack of equal-energy contours in Fig. 3(e) show that the Dirac SSs has a cone-like feature, which is rather isotropic. Compared with the experimental results, the Dirac point in the calculated SSs in Figs. 1(h) and 1(i) is closer to the conduction bands. The discrepancy is attributed to the difference in the surface conditions, which can change the band dispersions of SSs but do not change the topological properties.

The observation of a single Dirac cone across the bulk band gap is consistent with our theoretical predication that EuSn$_2$As$_2$ is a strong TI in the PM phase. As discussed above, the (001) Dirac SSs either open an energy gap or shift slightly off the $\bar{\Gamma}$ point in the AF phase. Within our resolution, we did not observe any significant changes in the Dirac SSs between the PM and AF phases. According to our calculations, the effect of the magnetic order on the Dirac SSs is very limited. The main reason is that the coupling between the electronic states involved in the magnetic order and those in the nontrivial topology are too weak. The magnetic moments derive from the Eu 4*f* states, which are very localized and lie at 1.7 eV below $E_F$. The nontrivial topology is caused by the inversion of the Sn 5*p* and As 4*p* states near $E_F$. In the AF phase, the Eu 4*f* magnetic moments form a long-range order, which breaks the time-reversal symmetry, allowing the surface Dirac bands to hybridize and develop an energy gap at $\bar{\Gamma}$. The magnitude of hybridization depends on the hopping probability between the Eu 4$f$ states and the $p$ orbitals related to the nontrivial topology. Since the topological electronic states has little contribution from the Eu 4$f$ states, the effects of magnetic order on the Dirac SSs are very limited.

We have revealed that EuSn$_2$As$_2$ is a magnetic TI with no observable gap in the Dirac SSs. The behavior seems to be significantly different from the large gap of tens of meV predicted in another widely studied magnetic TI family MnBi$_{2n}$Te$_{3n+1}$ ($n$ =1 and 2) [16,18,33]. The Mn 3$d$ magnetic moments in MnBi$_{2n}$Te$_{3n+1}$ form a long-range $A$-type AF order below $T_N$, which breaks the time-reversal symmetry while the inversion symmetry is preserved. The theoretical calculations have predicted that MnBi$_{2n}$Te$_{3n+1}$ is also axion insulators in the AF phase, which is analogous to EuSn$_2$As$_2$ except for the large gap in the (001) SSs of MnBi$_{2n}$Te$_{3n+1}$ in the calculations [16,28,33]. To clarify the effects of the Mn 3$d$ orbitals, especially on the topological SSs, we have performed precise ARPES measurements on the (001) surface of MnBi$_{2n}$Te$_{3n+1}$ ($n$ =1 and 2).

Figure 4 shows our synchrotron and laser ARPES data of MnBi$_2$Te$_4$. Compared with the previous ARPES results that exhibit a large gap of ~ 100 meV [21,25-27], our data reveal extra electronic states within the gap. In order to clarify the band dispersions of the in-gap states in the synchrotron ARPES data in Figs. 4(a) and 4(c), we performed second derivative of the ARPES data with respect to energy. In the intensity maps of second derivative in Figs. 4(b) and 4(d), two bands linearly cross, forming a Dirac point at -0.28 eV. The laser ARPES data in Figs. 4(f)-4(h) more clearly illustrate the Dirac-like band dispersions of the SSs. Moreover, we observed splitting in the conduction and valence bands below $T_N$, which can be attributed to band folding along $k_z$ and band reconstruction as the long-range AF order with an $A$-type magnetic structure leads to the doubling of the period along the $c$ axis. By carefully examining the band dispersions near the Dirac point in Figs. 4(i) and 4(j), we reveal that the upper and lower branches of the Dirac SSs do not touch, corresponding to the double-peak structure of the energy distribution curves (EDCs) at $\bar{\Gamma}$ in Fig. 4(k) and their second-derivative spectra in Fig. 4(l).

Note that the gap exists in both PM and AF phases of MnBi$_2$Te$_4$ ($T_N$ = 24 K). In Fig. 4(l), the second-derivative spectra of the EDCs at $\bar{\Gamma}$ determines that the gap is 12 meV at 40 K and 13.5 meV at 8 K. This difference is within the energy resolution ($\Delta E$ = 4.5 meV) in the experiments, indicating that the AF order has quite small effect on the Dirac SSs. This is in contrast to the theoretical calculations that have proposed that the Dirac SSs open an energy gap of tens of meV when $MnBi_2Te_4$ undergoes an AF transition into the AF-*c* phase [19-21,23,40]. The AF-*c* phase in $MnBi_2Te_4$ has been confirmed by neutron diffraction experiments [35].

In $MnBi_2Te_4$, the magnetic moments derive from the Mn 3*d* states, and the nontrivial topology is caused by the inversion of the Bi 6*p* and Te 5*p* states near the bulk band gap. In order to clarify the contribution of magnetic Mn 3*d* states to the topological electronic states, we performed resonance photoemission spectroscopy measurements at the Mn 3*p*−3*d* absorption edge. The difference between the on- and off-resonance spectra in Fig. 4(e) reveal that the Mn 3*d* states are mainly located at ~ 4 eV below $E_F$. In addition, the feature at ~ 1 eV below $E_F$ originates from the hybridization of the Mn 3*d* and Te 5*p* states. However, the component of Mn 3*d* states is negligible in the energy range within 0.6 eV below $E_F$, where the nontrivial topology arises. This indicates that the hybridization of the Mn 3*d* states and the *p* orbitals related to the nontrivial topology is weak. The theoretical calculations may overestimate the effects of the magnetic order on the Dirac SSs even though the Mn 3*d* states are not as localized as the Eu 4*f* states. The inconsistency between experiment and theory calls for further theoretical investigation.

We also revealed the existence of Dirac SSs within the large gap previously observed in $MnBi_4Te_7$ [28,29], as shown in Figs. 5. In the single crystal X-ray diffraction data in Fig. 5(a), all the peaks can be indexed by the (00*l*) reflections of $MnBi_4Te_7$ with *c* = 23.811(2) Å, which is consistent with the previous results [36]. The magnetic susceptibility $\chi(T)$ in Fig. 5(b) with *H*//*c* shows a cusp at 12.6 K, which is an indication of an AF transition consistent with the previous studies [28,29,43]. The isothermal magnetization *M*(*H*) with *H*//*c* at 2 K in Fig. 5(c) exhibits a spin-flip transition, indicating that the magnetic moments are along the *c* axis in the AF phase.

Figures 5(d)-5(f) show a Dirac-like band crossing at -0.28 eV at the $\bar{\Gamma}$ point within the bulk band gap. In Figs. 5(g) and 5(h), one can hardly recognize a gap at the Dirac point. The EDCs at $\bar{\Gamma}$ in Fig. 5(i) and their second-derivative spectra in Fig. 5(j) appear

a single-peak structure. This indicates that the gap is dramatically reduced from MnBi$_2$Te$_4$ to MnBi$_4$Te$_7$. A comparative study of them may help to clarify the origin of the gap in the PM phase. The AF ordering in MnBi$_4$Te$_7$ ($T_N$ = 12.6 K) has little effect on the Dirac SS, indicating weak coupling between the magnetic Mn 3$d$ states and the topological electronic states, which is similar to the case in MnBi$_2$Te$_4$.

It is expected on the (001) cleaved surface of MnBi$_4$Te$_7$ that there should be two kinds of terminations [MnBi$_2$Te$_4$] and [Bi$_2$Te$_3$], which have different band dispersions of the surface states. However, we only observe one set of Dirac bands in the bulk band gap using the 7-eV laser. Our scanning tunneling microscopy measurements (not shown) confirm the existence of two different terminations on one cleaved surface. The sizes of each termination are typically less than 1×1 μm$^2$, which are much smaller than the spot size of the laser (~ 50×50 μm$^2$). The absence of the other set of surface Dirac bands in our data is probably due to the matrix element effects. Systematic photon-energy-dependent measurements may reveal the other set of surface Dirac bands.

We have observed the Dirac SSs across the bulk band gap of EuSn$_2$As$_2$ and MnBi$_{2n}$Te$_{3n+1}$ ($n$ = 1 and 2), demonstrating their nontrivial topology. The Dirac SSs have almost no change when the long-range AF order develops in these materials, which can be attributed to weak coupling between the local magnetic moments and the topological electronic states. It is highly desirable to find the intrinsic magnetic TIs in that the topological electronic states are heavily involved in the magnetic order, which may be critical to realize topological quantum phenomena at higher temperatures.

*Note added*: We become aware of similar studies in Ref. [44-46] and the updated version of Ref. [26] showing the Dirac SSs in MnBi$_2$Te$_4$ when finalizing our paper and during the review process.

**Method**

**Sample synthesis**

Single crystals of EuSn$_2$As$_2$ were grown by the Sn flux method at the Institute of Physics, Chinese Academy of Sciences. The high-purity Eu (rod), Sn (shot), and As (lump) were put into corundum crucibles and sealed into quartz tubes with a ratio of Eu:As:Sn = 1:3:20. The tubes were heated to 1000 °C at the rate of 100 °C/h and held there for 12 h, then cooled to 750 °C at a rate of 2 °C/h. The flux was removed by centrifugation, and shiny crystals were obtained.

Both single crystals of MnBi$_2$Te$_4$ and MnBi$_4$Te$_7$ were grown by the self-flux method at Renmin University of China. The high-purity Mn (piece), Bi (shot) and Te (shot) were put into corundum crucibles and sealed into quartz tubes with a ratio of Mn:Bi:Te = 1:11.7:18.55 (MnTe:Bi$_2$Te$_3$ = 1:5.85) and Mn:Bi:Te = 0.8:11.7:18.35 (MnTe:Bi$_2$Te$_3$ = 0.8:5.85), respectively. The tube was heated to 950 °C at a rate of 40 °C/h and held there for 12 h, then cooled to 585 °C at a rate of 10 °C/h. The flux was removed by centrifugation, and shiny crystals were obtained.

**Band structure calculations**

EuSn$_2$As$_2$ is crystallized in a rhombohedral lattice with space group of *R*-3*m* (No. 166). The experimental lattice constants *a* = *b* = 4.2071 Å and *c* = 26.463 Å were adopted in our first-principles calculations. The Eu, Sn and As atoms are located at the Wyckoff position 3*a* (0, 0, 0), 6*c* (0, 0, 0.20963) and 6*c* (0, 0, 0.40624), respectively. The Vienna Ab initio Simulation Package (VASP) with the generalized gradient approximation–Perdew, Burke and Ernzerhof (PBE–GGA)-type exchange correlation potential was employed and the BZ sampling was performed by using *k* grids with an 11×11×3 mesh in self-consistent calculations. To match the energy position of Eu 4*f* bands in experiments, the Hubbard *U* parameter of the 4*f* electrons was taken as 5 eV in the GGA+U calculations. We have generated the maximally localized Wannier functions for the 5*s* and 5*p* orbitals on Sn and the 4*p* orbitals on As using the WANNIER90 package. The surface states calculations were performed using the Green's function method based on the Wannier Tools package.

**Synchrotron and laser angle-resolved photoemission spectroscopy**

Synchrotron ARPES measurements on EuSn$_2$As$_2$ were performed at the "CASSIOPEE" beamline, SOLEIL, France, with a Scienta R4000 analyzer, and the "dreamline" beamline at the Shanghai Synchrotron Radiation Facility (SSRF) with a Scienta DA30 analyzer. Synchrotron ARPES measurements on $MnBi_2Te_4$ were performed at the 03U beamline at the Shanghai Synchrotron Radiation Facility (SSRF) and at the 13U beamline at the National Synchrotron Radiation Laboratory at Hefei with a Scienta DA30 analyzer. High-resolution ARPES measurements on $MnBi_2Te_4$ and $MnBi_4Te_7$ were performed using the 7-eV laser ARPES at the Institute of Solid Physics, University of Tokyo with a Scienta R4000 analyzer.

**Time-resolved angle-resolved photoemission spectroscopy**

The tr-ARPES experiments were performed at Shanghai Jiao Tong University. In tr-ARPES measurements, infrared photon pulses with wavelength centered at 700 nm (1.77 eV) and pulse length of 30 fs were used to excite the sample, and the non-equilibrium states were probed by ultraviolet pulses at 205 nm (6.05 eV). Photoelectrons were collected by a Scienta DA30L-8000R analyzer. The overall time resolution and energy resolution are 130 fs and 19 meV, respectively [47]. Sample was cleaved at a pressure better than $3\times10^{-11}$ torr at 4 K.

**Acknowledgements**

T.Q. thanks Kun Jiang for discussions. We thank Quanxin Hu, Zhenyu Yuan and Jinpeng Xu for scanning tunneling microscopy measurements on $MnBi_4Te_7$. This work was supported by the Ministry of Science and Technology of China (2016YFA0300600, 2018YFA0305700, 2016YFA0300500, 2017YFA0302901, and 2016YFA0401000), the National Natural Science Foundation of China (11622435, U1832202, 11674369, 11674224, 11774399, 11574394, 11774423, 11822412, 11227907 and 11888101), the Chinese Academy of Sciences (QYZDB-SSW-SLH043 and XDB28000000), the CAS Pioneer "Hundred Talents Program" (type C), the Fundamental Research Funds for the Central Universities, the Research Funds of Renmin University of China (15XNLQ07, 18XNLG14 and 19XNLG17), the Beijing Municipal Science & Technology Commission (Z171100002017018, Z181100004218001 and Z181100004218005), the Beijing Natural Science Foundation (Z180008), the Science Challenge Project (TZ2016004), and the K. C. Wong Education Foundation (GJTD-2018-01). Laser-ARPES work was supported by the JSPS KAKENHI (JP18H01165 and JP19H00651) and Grant-in-Aid for JSPS Fellows (19F19030).

TABLE I. Numbers of occupied bands of odd parity at the eight TRI points in the PM and AF-*b*/AF-*c* phases of EuSn$_2$As$_2$, respectively.

| TRIM | (0, 0, 0) | (π, 0, 0) | (0, π, 0) | (π, π, 0) | (0, 0, π) | (π, 0, π) | (0, π, π) | (π, π, π) |
|---|---|---|---|---|---|---|---|---|
| PM | 100 | 102 | 102 | 102 | 102 | 102 | 102 | 102 |
| AF-*b*/*c* | 230 | 232 | 232 | 232 | 225 | 225 | 225 | 225 |

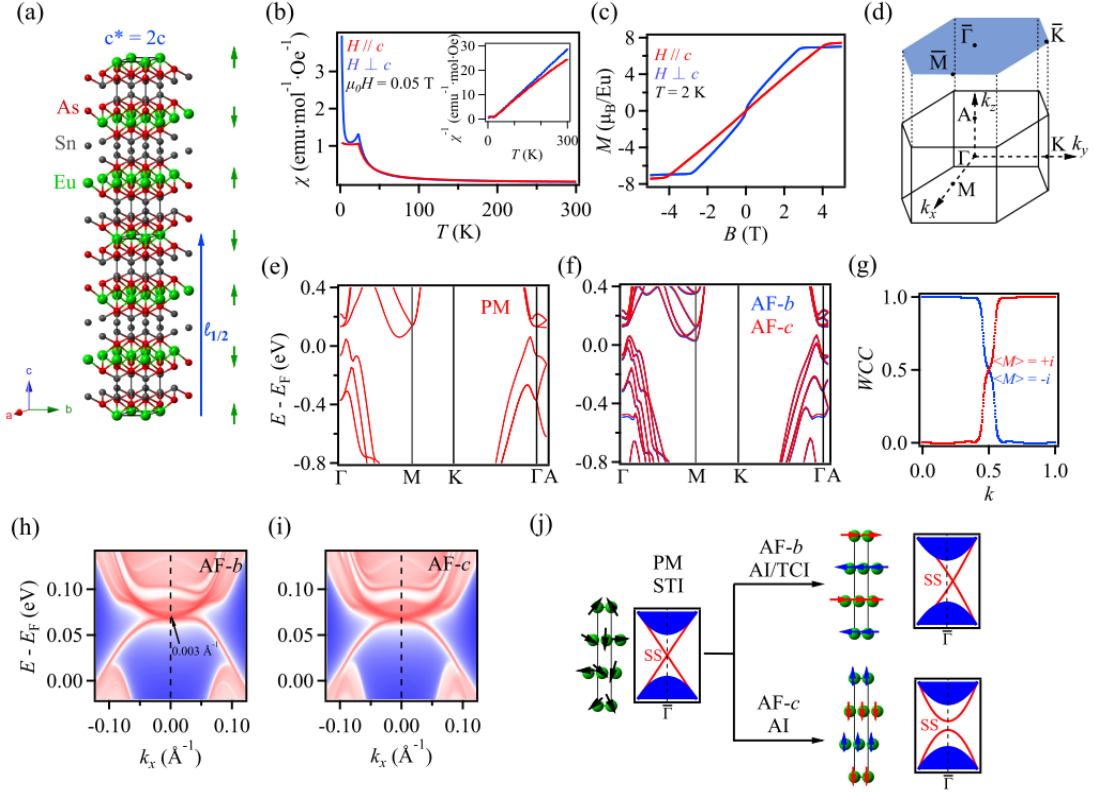

FIG. 1. Magnetic and topological properties of EuSn$_2$As$_2$. (a) Crystal structure of EuSn$_2$As$_2$. The green arrows represent the magnetic moments of Eu atoms, forming a doubled magnetic unit cell $c^* = 2c$ in the AF phase. The blue arrow represents the half translation operator $\ell_{1/2}$ of the magnetic unit cell along the $c$ axis. (b) Temperature dependence of magnetic susceptibility $\chi(T)$ with magnetic field $\mu_0 H = 0.05$ T parallel (red) and perpendicular (blue) to the $c$ axis. The inset shows the inverse susceptibility. (c) Field-dependent magnetization at 2 K with magnetic fields parallel (red) or perpendicular (blue) to the $c$ axis. (d) Bulk BZ and (001) surface BZ. $k_y$ and $k_z$ are along the $b$ and $c$ axes, respectively. (e),(f) Calculated bulk band structures of EuSn$_2$As$_2$ in the PM and AF-$b$/AF-$c$ phases along high-symmetry lines with spin-orbit coupling. (g) Wannier charge centers (WCC) calculated for the occupied bands with mirror eigenvalue $\langle M\rangle = +i$ ($-i$) in the mirror plane $k_y = 0$. (h),(i) Calculated band dispersions of the (001) SSs and projected bulk states along $\bar{\Gamma} - \bar{M}$ near the bulk band gap in the AF-$b$ and AF-$c$ phases. (j) Schematic diagram of the topological states of EuSn$_2$As$_2$ in three prototypical magnetic phases (PM, AF-$b$, AF-$c$). STI, TCI, and AI are the abbreviations of strong topological insulator, topological crystalline insulator, and axion insulator, respectively.

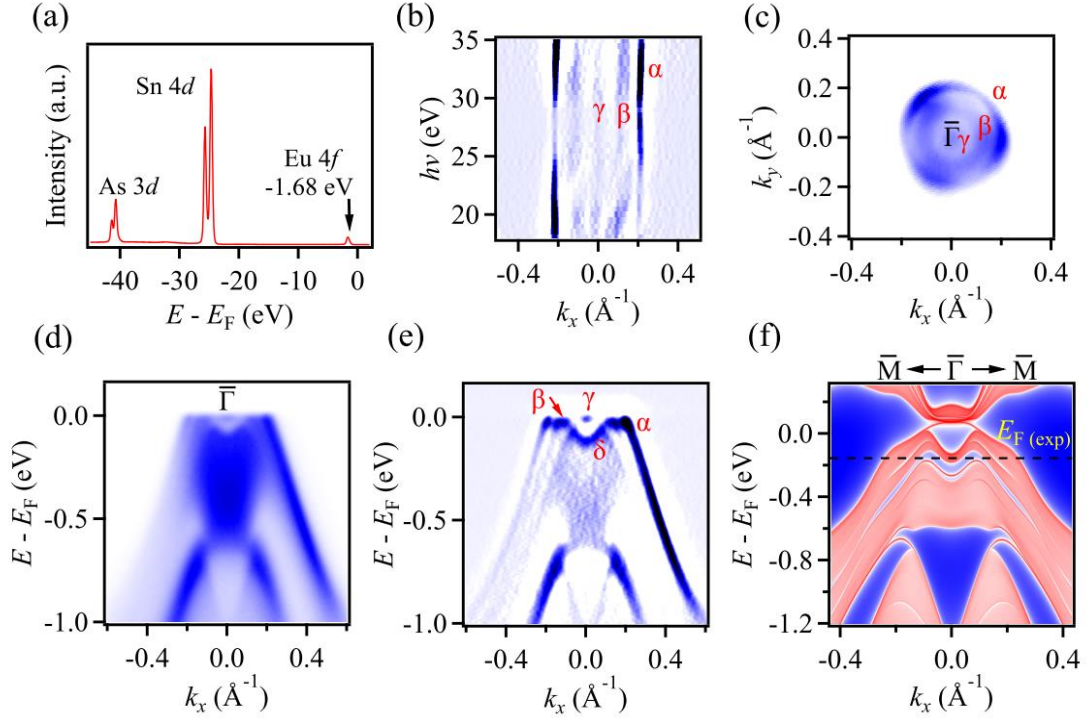

FIG. 2. Electronic structures below $E_F$ of EuSn$_2$As$_2$. (a) Core-level photoemission spectrum showing the characteristic peaks of As 3$d$, Sn 4$d$, and Eu 4$f$ orbitals. (b) Curvature intensity map of the ARPES data at $E_F$ along $\bar{\Gamma} - \bar{M}$ taken in a range of $h\nu$ from 18 to 35 eV. (c) ARPES intensity map at $E_F$ around $\bar{\Gamma}$. (d) ARPES intensity map along $\bar{\Gamma} - \bar{M}$. (e) Curvature intensity map of the data in (d). (f) Calculated band dispersions of the (001) SSs and projected bulk states along $\bar{\Gamma} - \bar{M}$. The ARPES data in (c) and (d) were collected at $h\nu$ = 29 eV. All ARPES data in Fig. 2 were collected at 50 K.

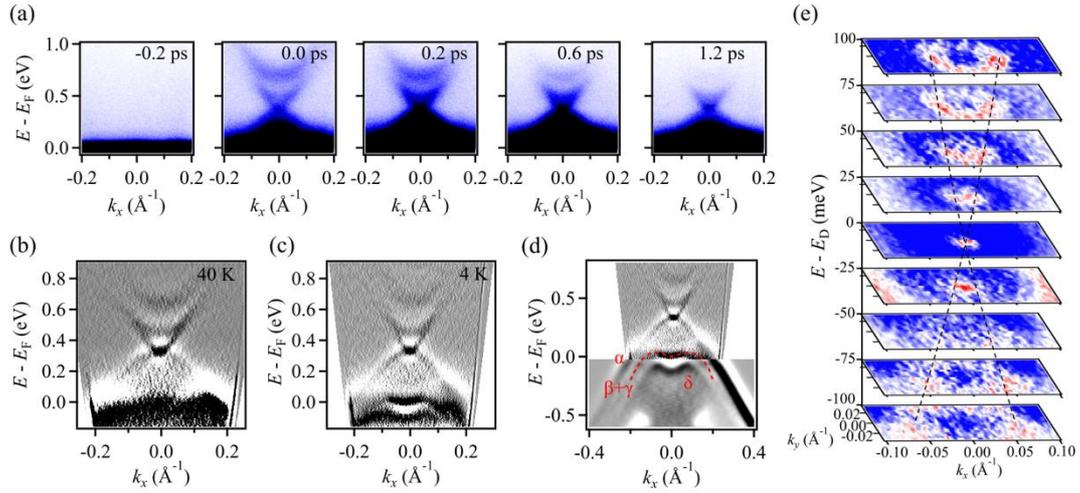

FIG. 3. Dirac SSs above $E_F$ on the (001) surface of EuSn$_2$As$_2$. (a) Snapshots of the tr-ARPES intensity along $\bar{\Gamma} - \bar{M}$ taken with the pump-probe method at different delay times. (b),(c) Curvature intensity maps of the pump-probe data at the delay time 0.2 ps taken at 40 K and 4 K. (d) Combination of the data in Figs. 3(c) and 2(e). (e) Stack of curvature intensity maps of the pump-probe data at different constant energies. The energy of the Dirac point ($E_D$) is set to zero.

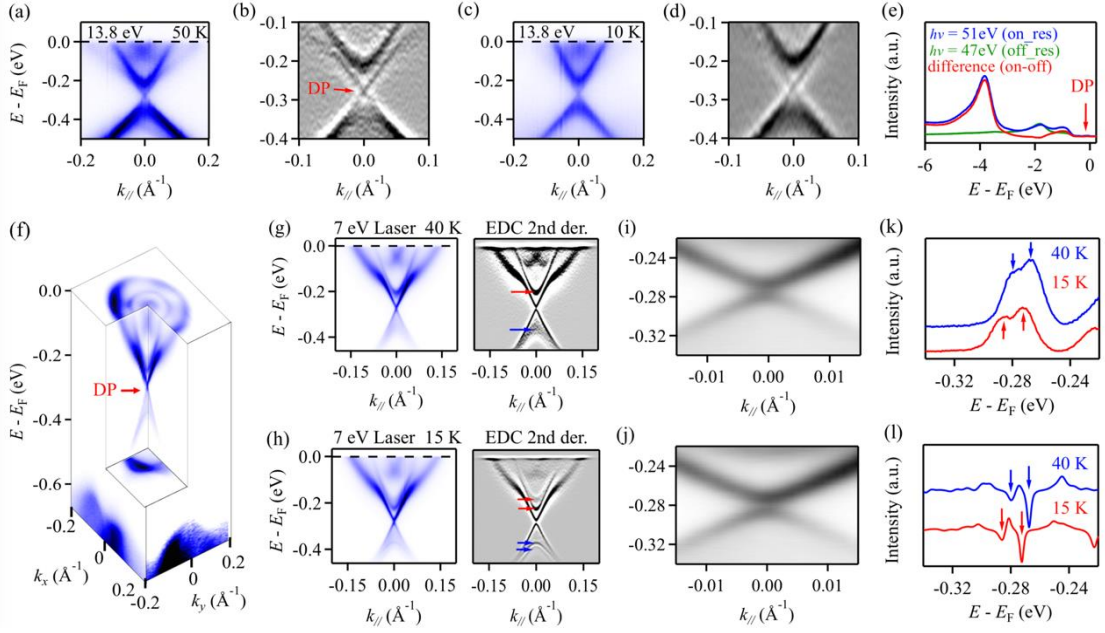

FIG. 4. Dirac SSs on the (001) surface of MnBi$_2$Te$_4$. (a),(c) ARPES intensity maps along the cut through $\bar{\Gamma}$ taken at $hv$ = 13.8 eV at 50 K and 10 K. (b),(d) Intensity maps of second derivative with respect to energy of the data near the bulk band gap in (a) and (c). DP is the abbreviation of Dirac point. (e) Resonant valence band spectra of MnBi$_2$Te$_4$ taken at the Mn 3$p$-3$d$ absorption edge. On and off-resonance spectra were obtained at $hv$ = 51 and 47 eV, respectively. (f) Three-dimensional plot of the ARPES data around $\bar{\Gamma}$ taken with the 7-eV laser. (g),(h) ARPES intensity maps (left) and their second derivative with respect to energy (right) along the cut through $\bar{\Gamma}$ taken with the 7-eV laser at 40 K and 15 K. The red and blue arrows indicate splitting of the conduction and valance bands, respectively, below $T_N$. (i),(j) Zoom-in ARPES intensity maps near the Dirac point in (g) and (h). (k) EDCs at $\bar{\Gamma}$ at 40 K and 15 K. (l) Second derivative of the EDCs in (k).

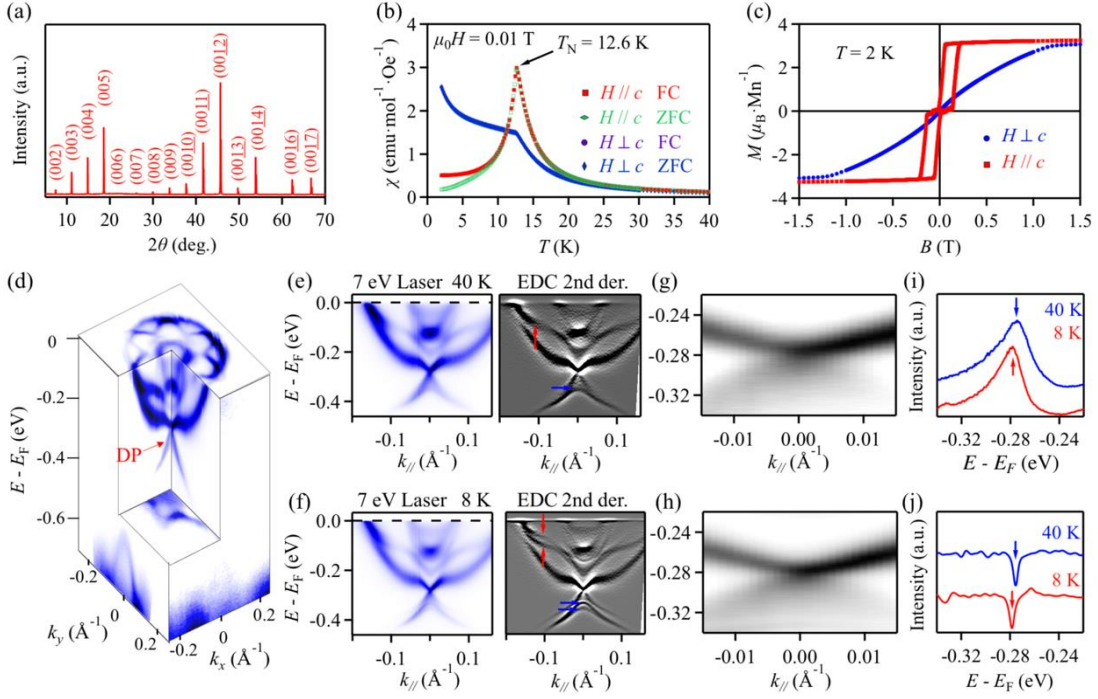

FIG. 5. Dirac SSs on the (001) surface of MnBi$_4$Te$_7$. (a) X-ray diffraction pattern on the (001) surface of MnBi$_4$Te$_7$ single crystal. (b) Temperature dependence of magnetic susceptibility $\chi(T)$ at magnetic field $\mu_0 H = 0.01$ T parallel and perpendicular to the $c$ axis under zero-field cooling (ZFC) and field cooling (FC). (c) Field-dependent magnetization $M(H)$ at 2 K with magnetic fields parallel (red) and perpendicular (blue) to the $c$ axis. (d) Three-dimensional intensity plot of the ARPES data around $\bar{\Gamma}$ measured with the 7-eV laser. (e),(f) ARPES intensity maps (left) and their second derivative with respect to energy (right) along the cut through $\bar{\Gamma}$ measured with the 7-eV laser at 40 K and 8 K. The red and blue arrows indicate splitting of the conduction and valance bands, respectively, across $T_N$. (g),(h) Zoom-in ARPES intensity maps near the Dirac point in (e) and (f). (i) EDCs at $\bar{\Gamma}$ measured at 40 and 8 K. (j) Second derivative of the EDCs in (i).